\documentclass[%
reprint,
 amsmath,amssymb,
 aps,
pra,
]{revtex4-1}

\usepackage{graphicx}
\usepackage{dcolumn}
\usepackage{bm}
\usepackage{braket}
\usepackage{color}
 
\begin{document}
\preprint{APS/123-QED}

\title{Realization of a timescale with an accurate optical lattice clock}

\author{C. Grebing}
 \email{christian.grebing@ptb.de}
\author{A. Al-Masoudi}
\author{S. D\"orscher}
\author{S. H\"afner}
\author{V. Gerginov}
\author{S. Weyers}
\author{B. Lipphardt}
\author{F. Riehle}
\author{U. Sterr}
\author{C. Lisdat}
\affiliation{%
 Physikalisch-Technische Bundesanstalt, Bundesallee 100, D-38116 Braunschweig, Germany
}%

\date{\today}

\begin{abstract}
Optical clocks are not only powerful tools for prime fundamental research, but are also deemed for the re-definition of the SI base unit second as they now surpass the performance of caesium atomic clocks in both accuracy and stability by more than an order of magnitude. However, an important obstacle in this transition has so far been the limited reliability of the 
optical clocks that made a continuous realization of a timescale 
impractical. In this paper, we demonstrate how this situation can be resolved and that a timescale based on an optical clock can be established that is superior to one based on even the best caesium fountain clocks. The paper also gives further proof of the international consistency of strontium lattice clocks on the $10^{-16}$ accuracy level, which is another prerequisite for a change in the definition of the second.
\end{abstract}

\maketitle


\section{Introduction}

The international system of units (SI) is the universal base for all measurements and thus constitutes the backbone of natural sciences and engineering. The definitions themselves and their realizations have been adapted over the years to profit from state-of-the-art research results. Important modifications are at hand for the electrical units and the mass, which will be defined through agreed-upon values of fundamental constants \cite{bip14}.


Due to their impressive progress \cite{cho10, let13, dub13, hin13, blo14, hun14, god14, ush15, lud15}, optical clocks surpass caesium fountain clocks, which currently realize the SI-second with lowest uncertainty, in terms of stability and of accuracy realizing an unperturbed atomic transition frequency. 
This has triggered discussions about the need for a re-definition of this unit \cite{rie15, mar13b}.
A first step has been the acknowledgement of suitable optical transitions in neutral atoms and ions as secondary representations of the SI unit ``second'' by the Commit\'e International des Poids et Mesures (CIPM) \cite{gil06b}. 

Given the outstanding role of the SI, it is of paramount importance that adaptations must be prepared carefully. 
Hence, changes will be acceptable only if \cite{rie15}
\begin{itemize}
	\item[$\circ$] an obvious candidate is identified,
	\item[$\circ$] the transition is smooth, and
	\item[$\circ$] the new approach is practical.
\end{itemize}
While the first point stimulates ongoing research, the second point has already been addressed satisfactorily for strontium optical lattice clocks \cite{let13, fal14} by measurements of the strontium clock transition frequency with today's best possible accuracy; the results in this paper confirm the previous measurement results. The latter prerequisite, however, has not been achieved so far for the candidate systems as existing optical clocks have generally been considered less reliable and thus less suitable for the actual implementation of a practical timescale, which continuously accumulates the atomic seconds.

\begin{figure}[b]
\centering
\includegraphics[width=\columnwidth]{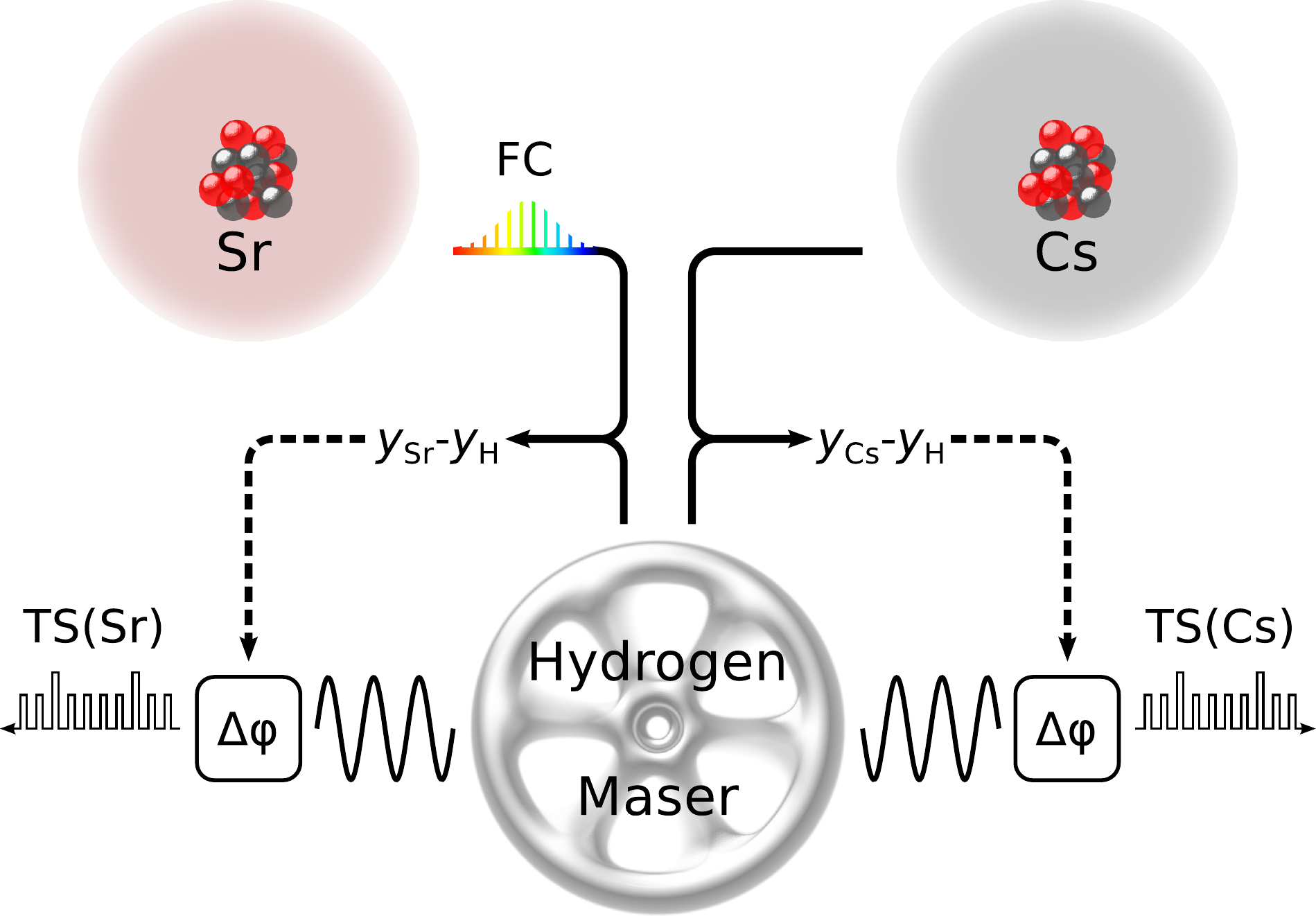}
\caption{Realization of a timescale TS from a microwave and an optical clock: The Cs clock transition frequency is compared against the maser flywheel frequency. The acquired offset $y_{\rm Cs}-y_{\rm H}$ is used to correct the classical timescale TS(Cs) generated from the maser utilizing a phase stepper ($\Delta\phi$). An equivalent scheme is applicable when referencing the timescale TS(Sr) to an optical frequency standard. For that purpose, the clock laser light is down-converted to the microwave regime using a femtosecond frequency comb (FC) before comparing against the flywheel. Moreover, both maser offsets can be analysed to yield the Sr clock frequency in SI units. 
}
\label{fig:network}
\end{figure}

Timescales provide us with coordinates for the position of events in time, much like a coordinate system does for the positioning in space. In particular, it needs to be realized without interruption to provide a continuous coordinate or to measure time intervals and to synchronize distant events. 

Today, the Universal Coordinated Timescale UTC is post-processed from a weighted monthly average of the time kept by some 400 microwave atomic clocks around the world that are intercompared via satellite links. To obtain time whenever needed, the time laboratories generate local timescales in real-time typically by steering the output frequency of a continuously running flywheel oscillator (Fig.~\ref{fig:network}), e.g. a hydrogen maser, on a monthly period to stay synchronized with the UTC clock ensemble \cite{par12a}. Some time laboratories synchronize their flywheel oscillators on an even shorter period to match the frequency of a local primary Cs clock \cite{bau12}.
 
The origin of the time error of such real-time timescales with respect to an ideal timescale is twofold: First, the accuracy of the local scale unit is limited by imperfections of the atomic reference(s). Even for the best caesium fountain clocks this error can integrate to about 1~ns after one month. Second, missing information from UTC or down-times of the local atomic reference clock
introduce a time error depending on the instability of the flywheel oscillator. 

Due to the averaging, UTC is more stable than most local timescales. Thus, in practice the quality of a local timescale is typically assessed by comparing to UTC, which, however, is not an ideal reference. Due to its construction, a deviation between the rate of UTC and the second of the local fountains can be as large as 0.1~ns/day, which accumulates to a time error of about 3~ns at the end of the month.



Optical clocks are expected to be capable of reducing the time error of a likewise realized local timescale by more than one order of magnitude (a projected clock uncertainty $\rm{of} \lesssim 1\times10^{-17}$ corresponds to a time $\rm{error}\lesssim 25$~ps after a month) -- provided the time error arising from their down-times is limited to an acceptable level -- and thus significantly improve the timescale's predictability. Together with improved time link technologies \cite{sli13}, a network of optical atomic clocks will allow the generation of a much more stable UTC and thus labelling events in time all over the world more precisely. This would be beneficial for global navigation systems, astro- and fundamental physics.

In this article, we demonstrate how 
PTB's strontium lattice clock is now able to maintain a local timescale with a time error of less than 200~ps compared to an ideal reference over about 25 days, or $7 \times 10^{-17}$ in fractional units. Over this period, the optical clock is allowed to operate with long interruptions (overall clock availability: $\approx46$~\%), and down-times are bridged by a hydrogen maser flywheel. The distortion of the scale unit due to the use of the flywheel while the clock is offline causes the dominant contribution to the achieved residual time error. Yet, this shows for the first time that already today optical frequency standards with limited availability can actually serve as atomic references to support a local timescale over extended periods, yielding a long-term performance better than of the ones referenced to the best current microwave clocks even if they are operated free of interruptions. 
%
Further improvements of the timescale based on an optical clock are at hand by ever higher availability of the optical clock. 

Moreover, the maser-flywheel-assisted Sr data together with data from one of our Cs fountains acquired in parallel enabled us to measure the Sr clock transition frequency with minimum uncertainty. So far this could only be achieved by operating the clocks for long times and/or by averaging over several of the best fountain clocks \cite{let13,fal14}.


\section{Optical clock operation} The strontium lattice clock was operated 
during two campaigns in October 2014 and June 2015
under conditions very similar to those of a previous campaign \cite{fal14}. 
A new interrogation laser system \cite{hae15a} improved the estimated optimum fractional frequency instability of the Sr clock, expressed as the Allan deviation $\sigma$, to below $\sigma_{\rm Sr}(\tau)=2 \times 10^{-16}$/$\sqrt{\tau}$ with averaging time $\tau$ \cite{alm15}. 

For best accuracy of the lattice clock, we reduced the duty cycle of the clock interrogation and thus reduced the power dissipation in the experimental apparatus that leads to thermal gradients. These gradients limit our knowledge of the ac- and dc-Stark line shift due to blackbody radiation (BBR) and thereby cause the largest uncertainty contribution to our lattice clock. With this reduced duty cycle, our clock still achieved an estimated instability of about $\sigma_{\rm Sr}(\tau)=5 \times 10^{-16}$/$\sqrt{\tau}$ during the measurement campaign. 
The clock's systematic uncertainty  $u_{\rm B}({\rm Sr})$ of $1.9 \times 10^{-17}$ has been evaluated similarly to \cite{fal14} and is discussed in detail in the supplement Section \ref{sec:suppl}. 
It is at least an order of magnitude smaller than the systematics of PTB's primary Cs fountain clocks CSF1 and CSF2 \cite{wey01,wey02,ger10,wey12}.

From October 6 through 15, 2014, the lattice clock was operated together with CSF2 ($u_{\rm B}({\rm CSF2})=3.1 \times 10^{-16}$) to measure the SI frequency of the Sr clock transition, which we will discuss first.
The frequencies of the Cs clock and the Sr lattice clock have been compared via femtosecond frequency combs (FC in Fig.~\ref{fig:network}). 
A continuously operated, high-performance hydrogen maser (VREMYA-CH VCH-1003M) used as a flywheel is connected to both clocks, e.g. for the generation of a timescale steered by either the Cs fountain (TS(Cs), \cite{bau12}) or the lattice clock (TS(Sr)).
During the measurement campaign, the lattice clock was operated over a total up-time of $T_{\rm Sr}=267\;000$~s (shown in Fig.~\ref{fig:stab} (b)), together with the almost continuously running fountain clock (availability $>98~\%$).

From this simultaneous comparison of the Cs and the Sr clock with the maser, the lattice clock's frequency can be calculated in the SI unit Hertz as realized by the Cs fountain clock. For white frequency noise, which is the dominating noise type in atomic frequency standards, the statistical clock uncertainty for a given averaging time $\tau$ is equal to the Allan deviation $u_{\rm clock}(\tau)=\sigma_{\rm clock}(\tau)$ \cite{lee10}. Thus, the fountain clock's instability $\sigma_{\rm CSF2}(\tau)=1.7 \times 10^{-13}$/$\sqrt{\tau}$ clearly dominates the statistical uncertainty of the measurement given by $u_{\rm A}=\sqrt{\sigma_{\rm Sr}^2(T_{\rm Sr}) + \sigma_{\rm CSF2}^2(T_{\rm Sr})}=3.3 \times 10^{-16}$.
 
However, using only the joint up-times of both clocks does not make any use of the information available from the maser: Its frequency is more stable than the fountain's for periods of up to $10^5$~s and can be measured accurately even during short up-times of the lattice clock. Therefore, it is reasonable to take the maser frequency value measured with the Sr clock as representative for longer intervals to improve $u_{\rm A}$.

In general, due to the maser instability there is a potential difference 
between the maser frequency averaged over time $T_{\rm Sr}$ and any chosen extended time $T_{\rm ext}$. The lack of knowledge about this potential difference 
is expressed as an additional uncertainty $u_{\rm ext}$, which, given the knowledge of the maser flywheel's properties, can be calculated.


\section{Data analysis} 
The uncertainty $u_{\rm ext}$ arises from the extrapolation of the mean flywheel frequency measured with our Sr clock over time $T_{\rm Sr}$ to the longer time $T_{\rm ext}$.
With the help of Parseval's theorem, $u_{\rm ext}$ can be expressed by the flywheel's spectrum of frequency fluctuations $S(f)$ and a weighting function $g(t)$ that describes the respective measurement intervals (see Section \ref{sec:suppl}).

\begin{figure}
\centering
\includegraphics[width=1\columnwidth]{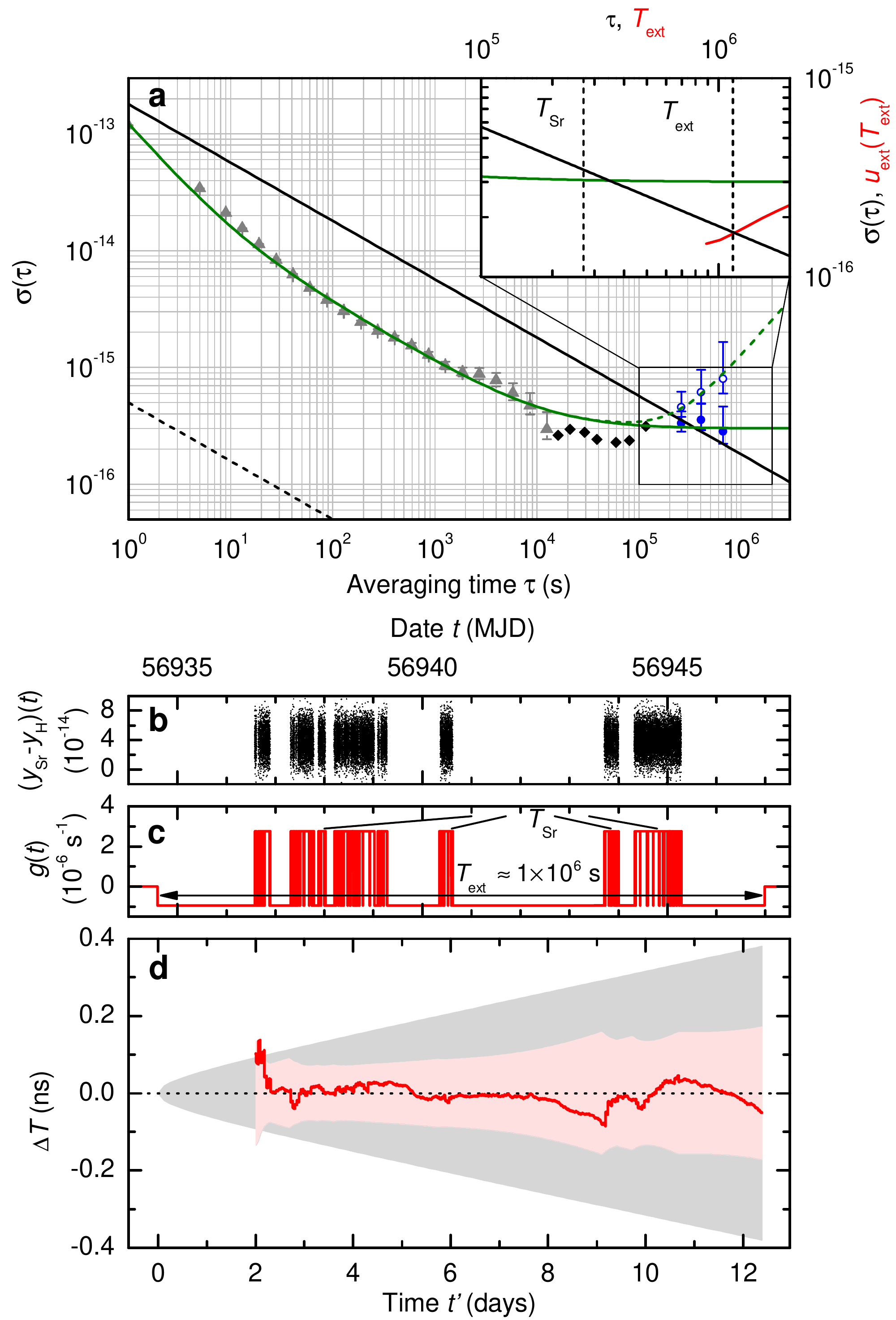}
\caption{Results achieved during measurement campaign 2014.\\
{\bf a}: Stability as represented by the Allan deviation $\sigma$ of the relevant oscillators. Solid black line -- fountain clock CSF2, dashed black line -- Sr lattice clock, data points -- measured maser stability (triangles: vs. lattice clock, diamonds: maser comparisons, empty dots: against fountain clocks, filled dots: ditto without linear drift), solid (dashed) green line -- noise model for the maser with (without) linear drift removed. The inset again shows the stabilities of the maser (green) and fountain clock (black). The red curve shows the additional uncertainty $u_{\rm ext}$ when the maser is used as a flywheel and data is extrapolated from an interval $T_{\rm Sr}=267\;000$~s to $T_{\rm ext}$. 
Dashed vertical lines indicate the direct and the optimum extrapolation measurement times.\\
{\bf b}: Frequency deviation between the nominal 100~MHz maser output and the Sr lattice clock averaged over 10~s assuming the Sr clock transition frequency equal to the recommendation of the CIPM for the secondary representation of the second \cite{cip13}, in total 267\;000~s. \\
{\bf c}: Weighting function used to derive the calibration uncertainty of the hydrogen maser's frequency with respect to the Sr lattice clock for an interval of $10^6$~s. \\
{\bf d}: 
Estimated 1-$\sigma$ time uncertainty range of TS(Sr) (red shaded area) and TS(Cs) (gray shaded area) including statistical and systematic contributions. The red solid line depicts the time error of a simulated timescale realization TS(Sr) with respect to an ideal reference. It is shown starting with the first corrected interval ($t'=0$ corresponds to the Modified Julian Date (MJD) $t\approx56934.6$).\\
}
\label{fig:stab}
\end{figure}

The maser spectrum $S(f)$ is extracted from different comparisons: the fast fluctuations and, thus, the stability at short averaging times $\tau$ are obtained from a direct comparison with the Sr clock. The Allan deviation of a long continuous data set is presented in Fig.~\ref{fig:stab} (a) (triangles). Mid- and low-frequency  information about the maser's frequency fluctuations are obtained from preceding maser -- maser comparisons (diamonds) and the measurements against the fountain clocks (dots, circles), respectively.
From these data, we derive a model for $S(f)$ that includes flicker frequency, white frequency, and flicker phase noise contributions. A linear maser drift was omitted here, because the analysis was designed to be drift-insensitive (see Section \ref{sec:suppl}). 


The given Sr up-times $T_{\rm Sr}$ (see Fig.~\ref{fig:stab} (b)) and the length and distribution of the extended time $T_{\rm ext}$, which we are free to choose, determine the weighting function $g(t)$ (see Fig.~\ref{fig:stab} (c) for the exemplary case of $T_{\rm ext}\approx10^6$~s) and, thus, together with $S(f)$ the corresponding extrapolation uncertainty $u_{\rm ext}$ (inset of Fig.~\ref{fig:stab} (a)). Due to the highly reliable Cs fountain it was not necessary to consider the influence of its down-times on $g(t)$.

For a flywheel-assisted Sr absolute frequency measurement we optimize $T_{\rm ext}$ such that the overall statistical measurement uncertainty $u_{\rm A} = \sqrt{\sigma_{\rm Sr}^2(T_{\rm Sr}) + \sigma_{\rm CSF2}^2(T_{\rm ext}) + u_{\rm ext}^2(T_{\rm ext})}$ is minimized.
As shown in the inset of Fig.~\ref{fig:stab} (a), the extrapolation uncertainty reaches the statistical uncertainty of the primary clock CSF2 at about $10^6$~s; further extension would degrade the combined statistical uncertainty $u_{\rm A}$. We can therefore enlarge the data set from 267\;000~s to about $10^6$~s. In consequence, the fractional systematic uncertainty of the primary fountain clock of $u_{\rm B}({\rm CSF2})=3.1 \times 10^{-16}$ becomes the largest contribution to the overall uncertainty of this frequency measurement of our strontium lattice clock. The result for the Sr clock transition frequency is 429\;228\;004\;229\;872.97(16)~Hz.

The evaluation procedure was applied accordingly to the later and longer measurement campaign with the Sr clock being operated from June 04 through 28 together with the same maser and both fountains: CSF1 ($u_{\rm B}({\rm CSF1})=7.0 \times 10^{-16}$) and CSF2 ($u_{\rm B}({\rm CSF2})=3.1 \times 10^{-16}$). The fountain clocks are considered independent and the results have been suitably averaged yielding the Sr clock transition frequency of 429\;228\;004\;229\;873.09(14)~Hz. Due to the significantly longer measurement time the overall statistical uncertainty is further improved. Yet, this does not affect the overall uncertainty considerably since it is governed by $u_{\rm B}({\rm CSF2})$. 

\begin{figure}
\centering
\includegraphics[width=1\columnwidth]{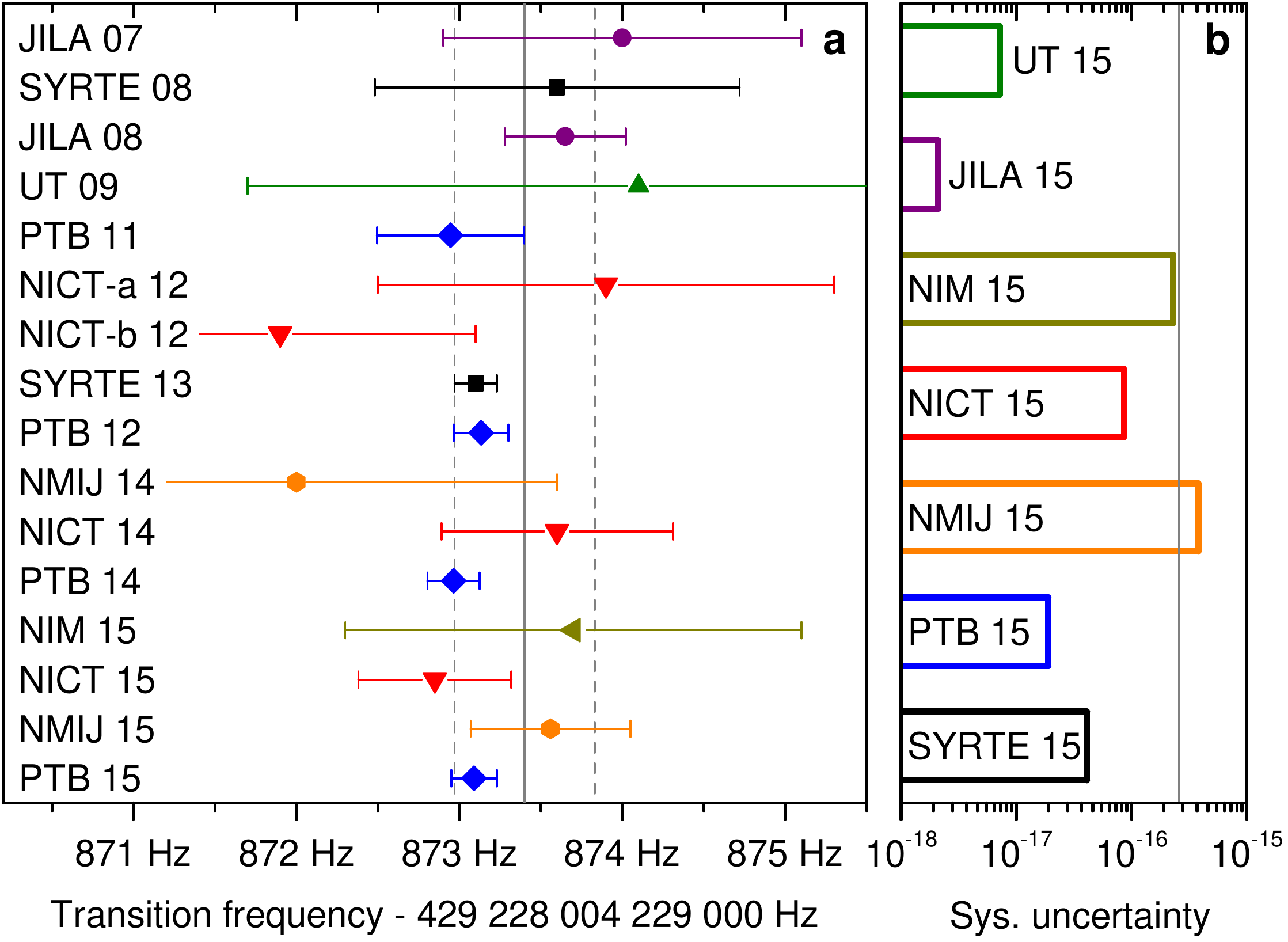}
\caption{
{\bf a}: Comparison of measured absolute frequencies of the 5s$^2$~$^1 S_0$~--~5s5p~$^3 P_0$ transition in $^{87}$Sr. The values have been obtained from various references \cite{boy07,bai08,cam08b,hon09,fal11,yam12,mat12,let13,fal14,aka14,hac14,lin15,hac15,tan15}. PTB~14 and PTB~15 have been obtained in this work. The vertical line indicates the frequency recommended by the CIPM in 2013 for the secondary representation of the second by Sr lattice clocks~\cite{cip13}, the dashed lines show the assigned uncertainty.\\
{\bf b}: Listing of the systematic uncertainty $u_{\rm B}$ of the best Sr clocks worldwide \cite{ush15,nic15,lin15,hac15,tan15,lis15}. The gray line indicates the Cs systematic uncertainty of the absolute frequency measurement with the smallest overall uncertainty so far \cite{let13}.
}
\label{fig:freq}
\end{figure}

Both measurements are in very good agreement with previous high-accuracy measurements \cite{let13,fal14} and provides a metrologically important confirmation that is necessary to build confidence in view of a re-definition of the SI-unit second. The excellent agreement between absolute frequency measurements of strontium lattice clocks in numerous institutes is shown in Fig.~\ref{fig:freq} (a). We want to emphasise that due to the assistance of the flywheel the achieved overall frequency measurement uncertainty in the 2015 campaign is very close to the lowest one ever achieved \cite{let13}. Likewise the frequency values deviate only in the low $10^{-17}$. Fig.~\ref{fig:freq} (b) points out that the systematic uncertainties of present-day Sr clocks worldwide range from similar to to well below the Cs systematics of the absolute frequency measurement with the lowest achieved uncertainty \cite{let13}.

\section{Realizing an optical timescale} 
The combination of the intermittently operated but accurate Sr clock and the reliable but less accurate maser can as well be utilized to establish a continuous high-performance timescale TS(Sr) by steering the maser output frequency.
This situation is not particular to optical clocks \cite{tan15,hac15,leu15}, but nowadays prevalent when timescales are established in metrology institutes or contributing to UTC \cite{lev12,par12a,yu07a,dou97}.

In general, the steering algorithm can be optimized with respect to a stable scale unit or a good long-term performance. Both design goals coincide for 100~\% clock up-time but they diverge the longer the clock down-times are that must be bridged with a less stable flywheel.
For example, when the Sr clock becomes available again after a long interruption, the flywheel instability may have lead to a large time error.
Thus, correcting the time error instantaneously will significantly distort the scale unit's stability while the stability for longer averaging times is improved.
For TS(Sr) we focus on a low statistical uncertainty for long averaging times.
In order to provide time whenever needed we choose a real-time implementation even though it 
can result in a degraded short-term stability compared to a post-processing approach.

The steering algorithm works as follows:
The average maser frequency is calculated regularly once per hour using the frequency information available up to that point from the Sr clock.
The average fractional frequency difference $\overline{y_{\rm H}-y_{\rm Sr}}$  is used as estimate for the fractional maser frequency $\overline{y}_{\rm H}$ over the full elapsed time $t'$. The corresponding time error $\Delta T'=-t'\cdot\overline{y}_{\rm H}$ is applied to the phase stepper in Fig.~\ref{fig:network} to correct the maser and yield TS(Sr). 

The evaluation of the extrapolation uncertainty discussed above can be used to quantify the crucial parameter of the timescale -- its predictability (or its uncertainty).
The main difference to the situation of the absolute frequency measurement is that now we are not free to chose the extended time interval $T_{\rm ext}$ since it is determined by the timescale's origin and the duration for which the timescale has to be provided. For convenience, we will discuss an optical timescale that covers the 12 days of our first flywheel-assisted absolute frequency measurement campaign.

The inset in Fig.~\ref{fig:stab} (a) shows the frequency uncertainty $u_{\rm ext}(T_{\rm ext})$ of an ensemble of masers -- all having the same instability like our maser -- whose frequencies are extrapolated from $T_{\rm Sr}$ to a continuous period $T_{\rm ext}$ (red line). The curve can be related to an accumulated timescale uncertainty via $u_{\rm ext}^{\rm TS}=u_{\rm ext}(T_{\rm ext})\cdot T_{\rm ext}$ when realizing a timescale by correcting such maser's frequency by intermittent measurements with our optical clock. To yield the overall optical timescale uncertainty, this contribution has to be added in quadrature to the accumulated time uncertainty of the Sr clock's scale unit, which is at least an order of magnitude smaller. 
The uncertainty of TS(Sr) is compared to that of a theoretical timescale TS(Cs) based on a Cs reference with 100~\% up-time (assuming the parameters of PTB's fountain clock CSF2). 

Fig.~\ref{fig:stab} (d) shows 1-$\sigma$-uncertainty bands of the timescales TS(Sr) and TS(Cs) as shaded areas including statistical and systematic contributions for the 2014 measurement campaign. In case of TS(Sr), the statistical uncertainty connected to the interrupted optical clock operation $u_{\rm ext}^{\rm TS}$ is calculated by the method described above, applying a weighting function that reflects the elapsed time and up-time of the optical clock. In this case, $u_{\rm ext}^{\rm TS}$ is not insensitive to a linear drift of the flywheel for all times and thus an uncertainty contribution considering this fact is added in quadrature. However, in a real operation scenario the drift is usually known, e.g. from long-term comparisons against UTC, and can easily be corrected.

For the considered 12-days period, the uncertainty of TS(Sr) is well below 200~ps at all times. While the overall uncertainty of TS(Cs) is dominated by the reference clock's systematics, that of TS(Sr) is governed by $u_{\rm ext}^{\rm TS}$ due to its limited availability of the Sr clock of about 27~\%. Thus, with increasing optical clock availability the uncertainty of TS(Sr) would be further reduced. Yet, even with the optical clock operation periods at hand, which can be considerably increased for the realization of a timescale, the uncertainty of TS(Sr) is clearly below that of the traditional timescale at all relevant times. Moreover, the uncertainty of TS(Cs) is increasing much faster. 

To illustrate a typical behavior of TS(Sr) during the measurement interval of interest we choose a numerical approach (see Section \ref{sec:suppl})
since an experimental characterization requires an equal or even better reference timescale.
A resulting TS(Sr) is given in Fig.~\ref{fig:stab} (d) indicating its high stability. 
The initial transient response at day 2 is an artefact of the initialization of the real-time timescale and could be reduced considerably in post-processing. Between day 6 and 9, when no data from the lattice clock is available, TS(Sr) starts to deviate according to the free-running maser instability, but once there is new optical data available ($t>9$~days) the time offset does not increase further and is even reduced due to a improved estimate of the past maser average frequency. 


\begin{figure}
\centering
\includegraphics[width=1\columnwidth]{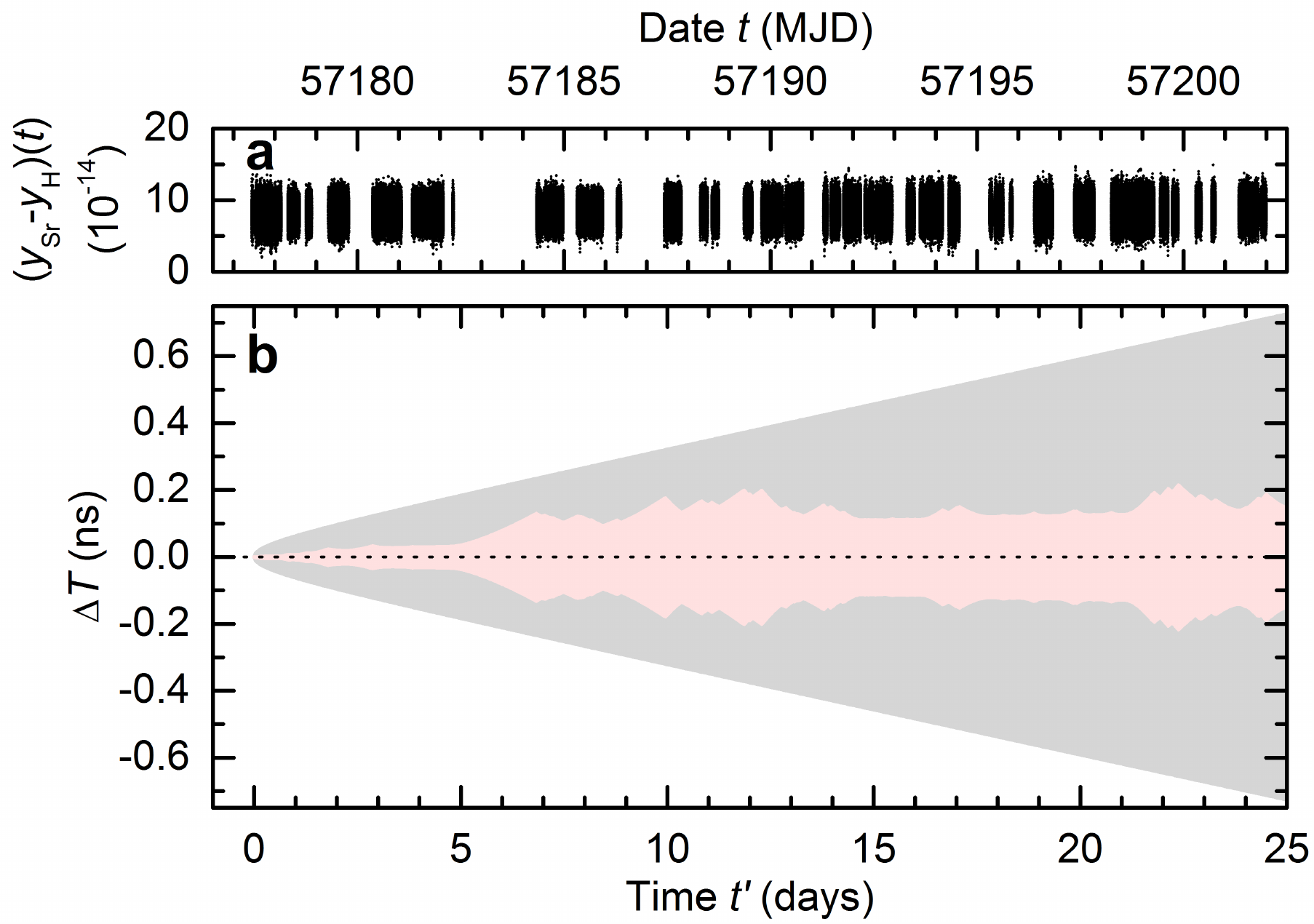}
\caption{Results achieved during measrurement campaign 2015.\\
{\bf a}: Frequency deviation between maser and the Sr clock.\\
{\bf b}: 1-$\sigma$ time uncertainty range of TS(Sr) (red) and TS(Cs) (gray) including statistical and systematic contributions.
}
\label{fig:timescale2015}
\end{figure}

The approach described above was also applied to the maser-assisted Sr data acquired in 2015 where the Sr-maser comparison was performed not only for a longer time ($T_{\rm Sr}\approx 922\;000$~s) but also with a much higher overall  availability of about 46~\% (see Fig.~\ref{fig:timescale2015} (a)). This allows providing a stable TS(Sr) over the extended period of 25 days, which is close to the typical 1-month reporting interval of UTC.
The uncertainty bands of TS(Sr) compared to TS(Cs) are depicted in (Fig.~\ref{fig:timescale2015} (b)). Still, the uncertainty of TS(Sr) is governed by
$u_{\rm ext}^{\rm TS}$. Yet, the higher optical clock availability enables a timescale uncertainty of $\lesssim 200$~ps for the whole 25-days period, which is a factor of 3.6 smaller than that of TS(Cs) that is controlled by an interruption-free CSF2. Thus, compared to the 2014 campaign no larger timescale uncertainty was accumulated even though the timescale was provided for an interval of twice the length.

The agreement of the flywheel-assisted strontium absolute frequency measurement presented above and the conventional measurements (Fig.~\ref{fig:freq} (a)) at the low $10^{-16}$ level can be considered a confirmation of our uncertainty evaluation (a frequency difference of $3\times10^{-16}$ corresponds to a time error of $\approx300$~ps (650~ps) over 12~days (25~days)).



\section{Conclusion}
\label{sec:meas}

This paper demonstrates that optical frequency standards and in particular strontium lattice clocks have reached a maturity such that they now can be used in combination with a high-quality commercially available flywheel oscillator to generate a timescale that shows beyond state-of-the-art performance. This is essential for actual optical clock applications and a redefinition of the SI-unit second.
The measurement campaign presented in Fig.~\ref{fig:stab}~(b) 
did not explicitly aim for maximum up-time and can therefore be regarded a lower limit of capability of our apparatus. Yet, Fig.~\ref{fig:stab} (d) shows that a time error of below 200~ps was achieved for a measurement interval of 12~days. 
In the second measurement campaign a higher optical clock availability was realized yielding an improved time uncertainty of $\lesssim 200$~ps over a 25-days interval,
which is close to the typical comparison duration to establish UTC. This is a remarkably low value in view of the deviations in the ns-range between the best timescales reported in the monthly Circular-T \cite{Circular_T_Cs}, which, however, includes an additional time link error of about $\lesssim1$~ns.

Concerning the flywheel oscillator, several choices are available besides the H-maser used here to achieve lower scale unit distortions for certain gap durations. Cryogenic sapphire microwave oscillators \cite{har12}, cryogenic optical reference resonators \cite{sto98,kes12a, hag14}, and lasers stabilized by the method of spectral hole burning \cite{tho11a} show excellent mid- and long-term stability and could serve this purpose. Even an optical clock that is optimized for reliability instead of accuracy can be envisaged. Ultimately, an ensemble of flywheels is conceivable providing lowest instability for all occurring gap durations.
 
Today's satellite-based comparison techniques can readily provide timescale comparisons with uncertainties below 1~ns \cite{pan10}. Thus, having averaged for a year or more, the well-established satellite network would allow the intercontinental comparison of flywheel-assisted optical clocks in the low $10^{-17}$ regime. Even further, ESA's future space mission Atomic Clock Ensemble in Space (ACES) will provide enhanced frequency comparison capabilities over intercontinental distances \cite{cac09}, such that flywheel-supported optical clock comparisons will no longer be limited by the satellite link between them. Thus, intercontinental clock comparisons with uncertainty in the $10^{-18}$ regime after averaging over a month is within reach.

The current optical clocks can be developed towards mostly autonomous systems that will increase the average availabilities \cite{bar12b}. For example, a clock availability of 70~\% would immediately improve $u_{\rm ext}^{\rm TS}$ further to below 60~ps over 30~days assuming our system and a worst-case-scenario of a single 7~h interruption per day, which corresponds to a frequency uncertainty of $2\times 10^{-17}$. 

\begin{acknowledgments}
We gratefully acknowledge Andreas Bauch for helpful discussions. This work was supported by the Centre of Quantum Engineering and Space-Time Reseach (QUEST), the German Research Foundation (DFG) through the RTG~1729 \lq Fundamentals and Applications of ultra-cold Matter\rq \ and the CRC 1128~geo-Q \lq Relativistic geodesy and gravimetry with quantum sensors\rq, and the project \lq International time-scales with Optical Clocks\rq \ (ITOC), which is part of the European Metrology Research Programme (EMRP). The EMRP is jointly funded by the EMRP participating countries within EURAMET and the European Union.
\end{acknowledgments}

\clearpage
\section{Supplementary material}
\label{sec:suppl}

{\bfseries Sr lattice clock operation} 
To interrogate the 698~nm ${^1S_0} - {^3P_0}$ clock transition of $^{87}$Sr, an atomic beam from an oven is Zeeman-slowed and loaded into a magneto-optical trap (MOT) operated on the strong 461~nm ${^1S_0} - {^1P_1}$ transition. After less than 300~ms loading, the atoms are transferred into a second-stage MOT operated on the 689~nm intercombination line, ${^1S_0} - {^3P_1}$, that allows for laser cooling of the atoms to a few microkelvin.

During the last cooling phase, the atoms are loaded into a nearly horizontally oriented 1D-optical lattice at the Stark-shift cancellation wavelength of strontium near 813~nm \cite{kat03}. In the lattice, the atoms are spin-polarized to either the $m_f = +9/2$ or $-9/2$ level of the ground state. The polarization of the sample is purified by a short $\pi$ pulse on the clock transition in a homogeneous magnetic bias field and subsequent removal of remaining ground state atoms.

After removing high-energy atoms from the lattice by temporarily reducing the lattice depth \cite{fal14}, the clock transition is interrogated in a bias field of about 25~$\mu$T.
The excitation probability is estimated from the atoms in ground and excited state after the interrogation. 
To derive an error signal free from linear Zeeman shift the $m_f = \pm 9/2$ transitions are sequentially interrogated at both points of maximum slope; thus four interrogations update the error signal. 
A single interrogation sequence lasts about 1~s, or, if a dead time is introduced to reduce heat load and thermal gradients, one interrogation every 2~s is performed.

Different from our previous set-up \cite{fal14}, the 429~THz clock laser system interrogating the $^{87}$Sr $^1S_0$ -- $^3P_0$ clock transition was replaced such that stabilization to the cryogenic silicon cavity \cite{kes12a} is no longer required \cite{hag13}. The new clock laser system uses a 48~cm long reference resonator made from ultra-low expansion glass \cite{hae15a}. It enables regular coherent interrogation of 640~ms duration, which leads to a resonance linewidth of about 1.4~Hz full-width at half-maximum.

{\bfseries Sr lattice clock uncertainty contributions}

\begin{table}
\centering
\begin{tabular}{lr@{}lr@{}l}
\hline
{\bf effect} 		&	\multicolumn{2}{c}{\bf correction} 		& \multicolumn{2}{c}{\bf uncertainty} \\ 
								&	\multicolumn{2}{c}{($10^{-17}$)} 			& \multicolumn{2}{c}{($10^{-17}$)} \\ 
\hline
\hline
BBR room 				&	492&.9 																&	1&.28\\
BBR oven 				&	0&.94																	&	0&.94\\
second-order Zeeman &	3&.6 															& 0&.15\\
cold collisions &	0 & 																	&	0&.08\\
background gas collisions& 0& 												&	0&.4\\
line pulling 		&	0 & 																	&	0&.01\\
lattice scalar/tensor &	$-0$&.7 															&	0&.9\\
lattice E2/M1 	&	0 & 																	&	0&.34\\
hyperpolarisability &	$-0$&.39 													&	0&.18\\
tunnelling 			&	0 & 																	&	0&.21\\
probe light 		&	0 & 																	&	0&.01\\
optical path length error &	0 & 												&	0&.01\\
servo error 		&	0 & 																	&	0&.17\\
DC Stark shift 	&	0	& 																	& 0&.03\\
\hline
{\bf total} 		&	496&.4 																&	1&.9\\
\hline
\end{tabular}
\caption{Corrections applied to the measured Sr lattice clock frequency and their uncertainties during the June 2015 frequency measurement given in fractional units. Details are given in the text.
}
\label{tab:unc}
\end{table}

\emph{BBR shift:} The Stark shift caused by BBR was determined experimentally \cite{mid11}, and validated theoretically \cite{saf13} as well as experimentally \cite{ush15}. Recently, the dynamic correction factor $\nu_{\rm dyn}$, reflecting the difference in frequency shift between BBR and a static electric field with equal rms amplitude, was refined \cite{nic15}. 
In our experiment, the uncertainty due to BBR is limited by the uncertainty of the representative temperature $T_{\rm rep}$. 
Temperature differences of 1.3~K at maximum measured between the warmest ($T_{\rm max}$) and coldest ($T_{\rm min}$) point of the apparatus lead under the assumption of a uniform probability distribution of $T_{\rm rep}$ to the most probable temperature of $(T_{\rm max} + T_{\rm min})/2 \approx 294$~K and an uncertainty of $1.28 \times 10^{-17}$ of the fractional frequency correction of $492.9 \times 10^{-17}$. In addition, radiation from the hot oven reaches the atoms and we estimate a corresponding correction of $9.4(9.4) \times 10^{-18}$ \cite{fal14}.

\emph{Zeeman shifts:} The linear Zeeman shift, as well as the vector light shift, are removed by stabilizing the interrogation laser to the mean frequency of the $m_F = \pm 9/2$ Zeeman components \cite{wes11}. Moreover, this scheme provides continuous monitoring of the total magnetic field experienced by the atoms and thus enables a characterization and correction of the second-order Zeeman shift. We find a shift of $-36(1.5) \times 10^{-18}$. The uncertainty is governed by the uncertainty of the magnetic field derived from the line splitting. The uncertainty of the correction coefficient  and the vector light shift can be neglected.

\emph{Collisions:} Cold collisions and collisions with the background gas may cause frequency shifts. We measured the cold collision shift by variation of the atom density in the lattice and found a shift of $0(8) \times 10^{-19}$. From the lattice lifetime of 4~s, background gas collisions lead to a shift of $0(4) \times 10^{-18}$ based on \cite{gib13}.

\emph{Line pulling:} Line pulling is induced by overlapping transitions caused by atoms populating different levels or by polarization imperfections of the excitation laser such that not only $\pi$- but also $\sigma$-transitions are driven simultaneously. Investigation of the preparation sequence and the laser polarization as seen by the atoms allows us to assign a negligible correction with small uncertainty to the line pulling effect of $0(1) \times 10^{-19}$.

\emph{Lattice light shifts:} By varying the lattice depth between the usual spectroscopy setting of 72~$E_r$ and 110~$E_r$, with the recoil energy of the lattice $E_r$, we find a Stark shift cancellation frequency of 368\;554\;471(3)~MHz for mutually parallel lattice polarization and magnetic field orientation. During the measurement campaign the lattice light frequency is typically slightly offset against the Stark cancellation frequency, which leads to a  shift of the clock frequency of $7(9) \times 10^{-18}$. Besides the determination of the Stark shift cancellation frequency under the given experimental conditions of lattice polarization and orientation,  possible shifts due to multi-photon and higher-order multi-pole excitations have to be accounted for \cite{wes11}. Using the coefficients of \cite{wes11} for E2/M1 transitions and two-photon induced light shifts (hyperpolarisability) \cite{let13}, these higher-order effects cause shifts of $0(3.4) \times 10^{-18}$ and $-3.9(1.8) \times 10^{-18}$, respectively.

\emph{Tunnelling:} Resonant tunnelling in the lattice causes energy bands with finite widths that can be populated non-uniformly, which in turn causes a tunnelling-induced frequency shift. Bandwidths and corresponding shifts can be reduced by a tilt or acceleration of the lattice such that tunnelling is no longer resonant \cite{lem05, sia08}. As we prepare the atoms almost exclusively in the vibrational ground state in longitudinal direction and operate in a deep lattice, even a small lattice tilt of $\theta=0.12(5)^\circ$ present in our set-up is sufficient to reduce the uncertainty due to tunnelling to $2.1 \times 10^{-18}$.

\emph{Probe light shift:} The highly advanced interrogation laser \cite{hae15a} enables long interrogation times of 640~ms. Therefore, the necessary intensity to drive a $\pi$-pulse and thus the light shift caused by the clock laser are very small. With the coefficient published in \cite{bai07}, we find less than $1 \times 10^{-19}$ of fractional shift, which we use as uncertainty. 

\emph{Optical path length:} Changes of the optical path length between atoms and the clock laser cause a first-order Doppler shift and therefore errors in the frequency measurement. Although continuous drifts due to, e.g., slow temperature changes may not be overly harmful \cite{fal12}, clock-cycle-synchronous vibrations or rf heating of the AOM may. Lattice clocks offer the unique opportunity to stabilize the optical path length from the interrogation laser to the atom position determined by the mirror that forms the standing wave of the 1D~lattice. Evaluation of the lock signal \cite{fal12} reveals an upper limit of path length-induced shifts of $1 \times 10^{-19}$.

\emph{Servo error:} In practice, all reference cavities used to build narrow bandwidth lasers exhibit a frequency drift due to ageing of the resonator materials and residual temperature variations. Any drift of the clock laser frequency causes a servo error, since frequency corrections are applied discontinuously. We reduce this effect by continuously sweeping the laser frequency with a rate that is updated from the atomic signal. Then, only non-linearities in the drift cause a servo error. As our new reference cavity shows an exceptionally linear drift, the fractional uncertainty is $1.7 \times 10^{-18}$.

\emph{DC Stark shift:} Static electric fields in the interrogation region, for instance produced by patch charges on isolating surfaces, can cause substantial frequency shifts \cite{fal12}. We have evaluated possible offset fields from the analysis of the observed Stark shift induced by intentionally applied electric fields. With an upper limit for the offset field of about 16~V/m, we expect a DC Stark shift of below $3 \times 10^{-19}$, which we take as uncertainty \cite{fal14}.

During the measurement campaign performed in October 2014 we observed an increased temperature gradient across the apparatus resulting in a larger uncertainty of the Stark shift due to BBR. This lead to a total clock uncertainty of $3\times 10^{-17}$.

{\bfseries Theory of data extrapolation}
For the analysis we use the normalized frequency fluctuations $y(t)$ defined as 
\begin{equation}
y(t) = \frac{\nu(t)-\nu_0}{\nu_0}
\end{equation}
with the frequency $\nu(t)$ and the nominal frequency $\nu_0$.
These fractional quantities allow to simply compare fluctuations between oscillators of different average frequencies with frequency ratio $\beta=\nu_0^{(a)}/\nu_0^{(b)}$.
\begin{equation}
 \frac{\nu^{(a)}(t)/\nu^{(b)}(t)}{\nu_0^{(a)}/\nu_0^{(b)}} -1 
\approx \frac{\nu^{(a)}(t)-\beta\nu^{(b)}(t)}{\nu^{(a)}_0} 
= y^{(a)}(t)-y^{(b)}(t).
\end{equation}
To express the uncertainty introduced by the extrapolation, we follow an approach similar to the one used to relate Allan deviations and the oscillator frequency noise spectrum (see e.g. \cite{daw07} and references therein).
Averaging the flywheel frequency $y(t)$ over one set of time intervals $\mathbb{T}_i$ with total duration $T_i$ can be written as
\begin{equation}
\overline{y}_{\mathbb{T}_i} = \int_{-\infty}^{\infty} y(t) g_{\mathbb{T}_i}(t) {\rm d}t,
\label{eq:udev}
\end{equation}
with the weighting function $g_{\mathbb{T}_i}(t)$: 
\begin{equation}
g_{\mathbb{T}_i}(t) = \left\{  \begin{array}{ll} 
				1/T_i 		& {\rm for \;} t\in \mathbb{T}_i \\
				0  							& {\rm elsewhere} \\
																	\end{array} \right..
\label{eq:sensfct}
\end{equation}
A similar approach is used for atomic clocks to describe the atoms' response to frequency fluctuations of the interrogation oscillator \cite{dic87}. Replacing the weighting function in Eq.~\ref{eq:udev} with the combined one  $g(t) = g_{\mathbb{T}_1}(t)-g_{\mathbb{T}_2}(t)$ yields the difference $\delta y_{\rm ext}=\overline{y}_{\mathbb{T}_1}-\overline{y}_{\mathbb{T}_2}$ of the mean flywheel frequencies averaged over the two sets of time intervals $\mathbb{T}_1$ and $\mathbb{T}_2$.
%

The uncertainty $u_{\rm ext}$ due to extrapolating the mean flywheel frequency from $\mathbb{T}_1$ to $\mathbb{T}_2$ is given by the standard deviation of this difference $\delta y_{\rm ext}$:
\begin{equation}
u_{\rm ext}^2 = \langle (\delta y_{\rm ext})^2 \rangle.
\label{eq:udev1}
\end{equation}
Parseval's theorem relates this variance to the single-sided power spectral density $S(f)$ of the flywheel oscillator's frequency fluctuations as:
\begin{equation}
u_{\rm ext}^2 = \int_0^\infty S(f) \left| G(f) \right|^2 {\rm d}f,
\label{eq:udev2}
\end{equation}
with $G(f)=\int_{-\infty}^{\infty} g(t)\exp(-2\pi{\rm i}\,ft){\rm d}t$ being the Fourier transform of the weighting function.
%
%
Note that the intervals $\mathbb{T}_1$ and $\mathbb{T}_2$ can deliberately be chosen symmetric about their common center-of-gravity. In this case,
a pure linear drift of $y(t)$ will lead to a coincidence of $\overline{y}_{\mathbb{T}_1}$ and $\overline{y}_{\mathbb{T}_2}$.
Thus, $\delta y_{\rm ext}$ and $u_{\rm ext}$ will not be affected by the linear drift of the flywheel oscillator.

{\bfseries Model of maser spectrum} Based on the observed maser frequency fluctuations after removing a linear frequency drift of $1.1 \times 10^{-16} \; /$day, the fractional instability of the maser $\sigma_{\rm H}(\tau)$ was modelled by three noise contributions that have been added in quadrature: A contribution of $1.18 \times 10^{-13}\cdot\tau^{-1}$ from white phase noise, a white frequency noise part of $3.5 \times 10^{-14}\cdot\tau^{-1/2}$ and a flicker floor of $3 \times 10^{-16}\cdot\tau^0$ \cite{par99, vor12}. These values were converted to corresponding spectral power densities of frequency fluctuations \cite{daw07} that were used in the calculation of $u_{\rm ext}$ using Eq. \ref{eq:udev2}.

{\bfseries Numerical timescale representation}
We numerically generate frequency data traces \cite{sta32} on a 10~s grid for the  maser flywheel and the Sr clock using the time intervals of the measurement campaign.
While the clock trace accounts for the statistical noise $\sigma_{\rm Sr}$, the maser trace reflects the noise and drift characteristics of our maser model (see above). 

This data is used to simulate a timescale realization TS(Sr) according to the steering algorithm.
At the beginning of the measurement interval, the time error $\Delta T$ between the timescale and an ideal one is set to zero. To attain a complete picture, one also has to account for the Sr clock's systematics $u_{\rm B}({\rm Sr})$. This is achieved by setting off the fractional clock frequency from its ideal value by $0.5\times u_{\rm B}({\rm Sr})$, which results in a minor time error of about 15~ps (20~ps) over the 12~days (25~days).

%

%



%

\end{document}